\documentclass[oneside,11pt]{article}

\usepackage[utf8]{inputenc}
\usepackage[T1]{fontenc}

\usepackage{amsthm}
\usepackage{amsmath}
\usepackage{amsfonts}
\usepackage{amssymb}

\usepackage{mathtools}

\usepackage[all]{nowidow}

\usepackage{graphicx} 
\usepackage{float}
\usepackage{booktabs}
\usepackage{multirow}
\usepackage{rotating}
\usepackage{array}

\usepackage{breakcites}

\usepackage{enumitem}

\usepackage[top=1.5cm,bottom=2cm,right=2.5cm,left=2.5cm]{geometry}
\usepackage{titling}

\usepackage{natbib}

\usepackage{ifplatform}

\ifwindows
\fi

\ifwindows
\usepackage{bbm}
\else
\iflinux
\usepackage{bbm}
\else
\usepackage{bbm}
\fi
\fi

\newcommand{\R}{\mathbb{R}}

\DeclareFontFamily{OT1}{pzc}{}
\DeclareFontShape{OT1}{pzc}{m}{it}{<-> s * [1.10] pzcmi7t}{}
\DeclareMathAlphabet{\mathpzc}{OT1}{pzc}{m}{it}

\usepackage[pdftex,final,bookmarksnumbered,bookmarksopen=false,breaklinks,colorlinks]{hyperref}
\hypersetup{
	final,
	bookmarksnumbered=true,
	bookmarksopen=true,
	bookmarksopenlevel=0,
	unicode=false,          
	pdftoolbar=true,        
	pdfmenubar=true,        
	pdffitwindow=false,     
	pdftitle={Bootstrap independence test for functional linear models},    
	pdfdisplaydoctitle=true, 
	pdftoolbar=true,
	pdfmenubar=true,
	pdflang={English},
	pdfauthor={Wenceslao Gonzalez-Manteiga, Gil Gonzalez-Rodriguez, Adela Martinez-Calvo, Eduardo Garcia-Portugues},     
	pdfsubject={arXiv paper},   
	pdfcreator={Eduardo Garcia-Portugues},   
	pdfproducer={Eduardo Garcia-Portugues}, 
	pdfkeywords={Bootstrap}{Bootstrap consistency}{Functional linear regression}{Functional principal components analysis}{Hypothesis test}, 
	pdfnewwindow=true,      
	breaklinks=true,
	hidelinks,       
	linkcolor=black,        
	citecolor=black,        
	filecolor=magenta,      
	urlcolor=cyan           
}

\newtheorem{rem}{Remark}

\newtheorem{lem}{Lemma}

\allowdisplaybreaks

\newcommand{\N}{\mathbb{N}}
\newtheorem{thm}{Theorem}
\newtheorem{cor}{Corollary}
\newtheorem{alg}{Algorithm}
\begin{document}

\title{Bootstrap independence test for functional linear models}
\setlength{\droptitle}{-1cm}
\predate{}%
\postdate{}%
\author{Wenceslao Gonz\'alez-Manteiga$^1$, Gil Gonz\'alez-Rodr\'iguez$^2$,\\ Adela Mart\'inez-Calvo$^{1,3}$, and Eduardo Garc\'ia-Portugu\'es$^{1}$}

\date{}

\footnotetext[1]{
	Department of Statistics and Operations Research, University of Santiago de Compostela (Spain).}
\footnotetext[2]{
	Department of Statistics and Operations Research and Mathematics Didactics, University of Oviedo (Spain).}
\footnotetext[3]{Corresponding author. e-mail: \href{mailto:adela.martinez@usc.es}{adela.martinez@usc.es}.}

\maketitle


\begin{abstract}
Functional data have been the subject of many research works over the last years. Functional regression is one of the most discussed issues. Specifically, significant advances have been made for functional linear regression models with scalar response. Let $({\cal H},\langle\cdot,\cdot\rangle)$ be a separable Hilbert space. We focus on the model $Y=\langle \Theta,X\rangle+b+\varepsilon$, where $Y$ and $\varepsilon$ are real random variables, $X$ is an ${\cal H}$-valued random element, and the model parameters $b$ and $\Theta$ are in $\R$ and ${\cal H}$, respectively. Furthermore, the error satisfies that $E(\varepsilon|X)=0$ and $E(\varepsilon^2|X)=\sigma^2<\infty$. A consistent bootstrap method to calibrate the distribution of statistics for testing $H_0: \Theta=0$ versus $H_1: \Theta\neq0$ is developed. The asymptotic theory, as well as a simulation study and a real data application illustrating the usefulness of our proposed bootstrap in practice, is presented.
\end{abstract}

\begin{flushleft}
\end{flushleft}

\section{Introduction}

Nowadays, {\em Functional Data Analysis\/} (FDA) has turned into one of the most interesting statistical fields. Particularly, functional regression models have been studied from a parametric point of view (see Ramsay and Silverman~(2002,~2005)), and from a non-parametric one (see Ferraty and Vieu~(2006)), being the most recent advances compiled on Ferraty and Romain~(2011). This work focuses on the parametric approach, specifically, on the {\em functional linear regression model with scalar response\/} that is described below.\\

Let $({\cal H},\langle \cdot,\cdot \rangle)$ be a separable Hilbert space, and let $\|\cdot\|$ be the norm associated with its inner product. Moreover, let $(\mathrm{\Omega}, \sigma, \mathrm{P})$ be a probability space and let us consider $(X,Y)$ a measurable mapping from $\mathrm{\Omega}$ to ${\cal H} \times \R$, that is, $X$ is an ${\cal H}$-valued random element whereas $Y$ is a real random variable. In this situation, let us assume that $(X,Y)$ verifies the following linear model with scalar response, 
\begin{equation}
Y = \langle \Theta, X \rangle + b + \varepsilon \label{eq.hlm}
\end{equation}
where $\Theta \in {\cal H}$ is a fixed functional model parameter, $b \in \R$ is the intercept term, and $\varepsilon$ is a real random variable such that $E(\varepsilon | X)=0$ and $E(\varepsilon^2|X)=\sigma^2<\infty$. Many authors have dealt with model \eqref{eq.hlm}, being the methods based on {\em Functional Principal Components Analysis\/} (FPCA) amongst the most popular ones to estimate the model parameters (see Cardot, Ferraty, and Sarda~(1999,~2003), Cai and Hall~(2006), Hall and Hosseini-Nasab~(2006), and Hall and Horowitz~(2007)).\\

The main aim of this work is to develop a consistent general bootstrap resampling approach to calibrate the distribution of statistics for testing the significance of the relationship between $X$ and $Y$, that is, for testing $H_0: \Theta=0$ versus $H_1: \Theta\neq0$, on the basis of a simple random sample $\{(X_i,Y_i)\}_{i=1}^n$ drawn from $(X,Y)$. The bootstrap techniques will become an alternative useful tool when the asymptotics of test statistics are unknown or when they are inaccurate due to small sample size.\\

Since its introduction by Efron~(1979), it is well-known that the bootstrap method results in a new distribution approximation which can be applied to a large number of situations, such as the calibration of pivotal quantities in the finite dimensional context (see Bickel and Freedman~(1981), and Singh~(1981)). As far as multivariate regression models are concerned, bootstrap validity for linear and non-parametric models was also stated in literature (see Freedman~(1981), and Cao-Abad~(1991)). Currently, the application of bootstrap to the functional field has been successfully started. For instance, Cuevas, Febrero, and Fraiman~(2006) have proposed bootstrap confidence bands for several functional estimators such as the sample and the trimmed functional means. In the regression context, Ferraty, Van~Keilegom, and Vieu~(2010), and Gonz\'alez-Manteiga and Mart\'inez-Calvo~(2011) have shown the validity of the bootstrap in the estimation of non-parametric functional regression and functional linear model, respectively, when the response is scalar. They have also proposed pointwise confidence intervals for the regression operator involved in each case. In addition, the asymptotic validity of a componentwise bootstrap procedure has been proved by Ferraty, Van~Keilegom, and Vieu~(2012) when a non-parametric regression is considered and both response and regressor are functional.\\

Bootstrap techniques can also be very helpful for testing purposes, since they can be used in order to approximate the distribution of the statistic under the null hypothesis $H_0$. For example, Cuevas, Febrero, and Fraiman~(2004) have developed a sort of parametric bootstrap to obtain quantiles for an ANOVA test, and Gonz\'alez-Rodr\'iguez, Colubi, and Gil~(2012) have proved the validity of a residual bootstrap in that context. Hall and Vial~(2006) and, more recently, Bathia, Yao, and Ziegelmann~(2010) have studied the finite dimensionality of functional data using a bootstrap approximation for independent and dependent data, respectively.\\

As was indicated previously, testing the lack of dependence between X and Y is our goal. This issue has stirred up a great interest during the last years due to its practical applications in the functional context. For instance, Kokoszka, Maslova, Sojka, and Zhu~(2008) proposed a test for lack of dependence in the functional linear model with functional response which was applied to magnetometer curves consisting of minute-by-minute records of the horizontal intensity of the magnetic field measured at observatories located at different latitude. The aim was to analyse if the high-latitude records had a linear effect on the mid- or low-latitude records. On the other hand, Cardot, Prchal, and Sarda~(2007) presented a statistical procedure to check if a real-valued covariate has an effect on a functional response in a nonparametric regression context, using this methodology for a study of atmospheric radiation. In this case, the dataset were radiation profiles curves measured at a random time and the authors tested if the radiation profiles changed along the time.\\

Regarding the regression model \eqref{eq.hlm}, testing the significance of the relationship between a functional covariate and a scalar response has been the subject of recent contributions, and asymptotic approaches for this problem can be found in Cardot, Ferraty, Mas, and Sarda~(2003) or Kokoszka, Maslova, Sojka, and Zhu~(2008). The methods presented in these two works are mainly based on the calibration of the statistics distribution by using asymptotic distribution approximations. In contrast, we propose a consistent bootstrap calibration in order to approximate the statistics distribution. For that, we firstly introduce in Section~2 some notation and basic concepts about the regression model \eqref{eq.hlm}, the asymptotic theory for the testing procedure, and the consistency of the bootstrap techniques that we propose. In Section~3, the bootstrap calibration is presented as an alternative to the asymptotic theory previously exposed. Then, Section~4 is devoted to the empirical results. A simulation study and a real data application allow us to show the performance of our bootstrap methodology in comparison with the asymptotic approach. Finally, some conclusions are summarized in Section 5.

\section{Asymptotic theory and bootstrap}

Let us consider the model \eqref{eq.hlm} given in the previous Section~1. In this framework, the regression function, denoted by $m$, is given by
\[m(x) = E(Y|X=x) = \langle \Theta, x \rangle + b \text{ for all } x \in {\cal H}.\]
The aim is to develop correct and consistent bootstrap techniques for testing
\begin{equation}
\left\{\begin{array}{ll} H_0: &\Theta=0 \\ H_1: &\Theta\neq 0 \end{array}\right. \label{eq.test}
\end{equation}
on the basis of a random sample $\{(X_i,Y_i)\}_{i=1}^n$ of independent and identically distributed random elements with the same distribution as $(X,Y)$. That is, our objective is to check whether $X$ and $Y$ are linearly independent ($H_0$) or not ($H_1$).\\

Next, we expose briefly some technical background required to develop the theoretical results presented throughout the section.

\subsection{Some background}

Riesz Representation Theorem ensures that the functional linear model with scalar response can be handled theoretically within the considered framework. Specifically, let ${\cal H}$ be the separable Hilbert space of square Lebesgue integrable functions on a given compact set $C \subset \R$, denoted by ${\cal L}^2(C,\lambda)$, with the usual inner product and the associated norm $\|\cdot\|$. The functional linear model with scalar response between a random function $X$ and a real random variable $Y$ is defined as
\begin{equation}
Y = \Phi(X) + \epsilon, \label{eq.flm}
\end{equation}
where $\Phi$ is a continuous linear operator (that is, $\Phi \in {\cal H}'$, being ${\cal H}'$ the dual space of ${\cal H}$ with norm $\|\cdot\|'$), and $\epsilon$ is a real random variable with finite variance and independent of $X$. In virtue of Riesz Representation Theorem ${\cal H}$ and ${\cal H}'$ are isometrically identified, in such a way that for any $\Phi \in {\cal H}'$ there exists a unique $\Theta \in {\cal H}$ so that $\|\Theta\|=\|\Phi\|'$ and $\Phi(h)=\langle \Theta,h\rangle$ for all $h \in {\cal H}$. Consequently, the model presented in equation \eqref{eq.flm} is just a particular case of the one considered in \eqref{eq.hlm}.\\

Previous works regarding functional linear models assume $b=0$ (see Cardot, Ferraty, Mas, and Sarda~(2003), and Kokoszka, Maslova, Sojka, and Zhu~(2008)). Of course, the intercept term can be embedded in the variable counterpart of the model as in the multivariate case as follows. Let ${\cal H}_e$ be the product space ${\cal H}\times \R$ with the corresponding inner product $\langle \cdot,\cdot\rangle_e$, and define $X'=(X,1)$ and $\Theta'=(\Theta,b) \in {\cal H}_e$. Then the model considered in \eqref{eq.hlm} can be rewritten as $Y = \langle \Theta', X' \rangle_e + \varepsilon$ (and consequently $X'$ cannot be assumed to be centered). Nevertheless, in the context of the linear independence test, the aim is to check if $\Theta=0$ or not, and this is not equivalent to checking whether $\Theta'=0$ or not. In addition, in practice the intercept term $b$ cannot be assumed to be equal to $0$. Thus, in order to avoid any kind of confusion, in this paper the intercept term $b$ has been written explicitly.\\

In the same way, in the above mentioned papers, the random element $X$ is assumed to be centered. Although, in many cases, the asymptotic distribution of the proposed statistics does not change if $\{X_i\}_{i=1}^n$ is replaced by the dependent sample $\{X_i-\overline{X}\}_{i=1}^n$, the situation with the bootstrap version of the statistics could be quite different. In fact, as it will be shown afterwards, different bootstrap statistics could be considered when this replacement is done. Hence, for the developments in this section, it will not be assumed that the $X$ variable is centered.

\subsection{Linear independence test}

Given a generic ${\cal H}$-valued random element $H$ such that $E(\|H\|^2)<\infty$, its associated covariance operator $\Gamma_H$ is defined as the operator $\Gamma_H:{\cal H} \rightarrow {\cal H}$ 
\[\Gamma_H(h)=E\left(\langle H-\mu_H,h\rangle(H-\mu_H)\right)=E\left(\langle H,h\rangle H\right)-\langle\mu_H,h\rangle\mu_H,\]
for all $h \in {\cal H}$, where $\mu_H \in {\cal H}$ denotes the expected value of $H$. From now on, it will be assumed that $E(\|X\|^2)<\infty$, and thus, as a consequence of H\"{o}lder's inequality, $E(Y^2)<\infty$. Whenever there is no possible confusion, $\Gamma_X$ will be abbreviated as $\Gamma$. It is well-known that $\Gamma$ is a nuclear and self-adjoint operator. In particular, it is a compact operator of trace class and thus, in virtue of the Spectral Theorem Decomposition, there is an orthonormal basis of $H$, $\{v_j\}_{j \in \N}$, consisting on eigenvectors of $\Gamma$ with corresponding eigenvalues $\{\lambda_j\}_{j \in \N}$, that is, $\Gamma(v_j)=\lambda_j v_j$ for all $j \in \N$. As usual, the eigenvalues are assumed to be arranged in decreasing order ($\lambda_1\geq \lambda_2\geq \ldots$). Since the operator $\Gamma$ is symmetric and non-negative definite, then the eigenvalues are non-negative.\\

In a similar way, let us consider the cross-covariance operator $\Delta: {\cal H} \rightarrow \R$ between $X$ and $Y$ given~by
\[\Delta(h)=E\left(\langle X-\mu_X,h\rangle(Y-\mu_Y)\right)=E\left(\langle X,h\rangle Y\right)-\langle\mu_X,h\rangle\mu_Y,\]
for all $h \in {\cal H}$, where $\mu_Y \in \R$ denotes the expected value of $Y$. Of course, $\Delta \in {\cal H}'$ and the following relation between the considered operators and the regression parameter $\Theta$ is satisfied
\begin{equation}
\Delta(\cdot)=\langle \Gamma(\cdot), \Theta\rangle. \label{eq.relation}
\end{equation}

The Hilbert space $\cal H$ can be expressed as the direct sum of the two orthogonal subspaces induced by the self-adjoint operator $\Gamma$: the kernel or null space of $\Gamma$, ${\cal N}(\Gamma)$, and the closure of the image or range of $\Gamma$, $\overline{{\cal R}(\Gamma)}$. Thus, $\Theta$ is determined uniquely by $\Theta=\Theta_1 + \Theta_2$ with $\Theta_1 \in {\cal N}(\Gamma)$ and $\Theta_2 \in \overline{{\cal R}(\Gamma)}$. As $\Theta_1 \in {\cal N}(\Gamma)$, it is easy to check that $Var(\langle X,\Theta_1\rangle)=0$ and, consequently, the model introduced in \eqref{eq.hlm} can be expressed as
\[Y = \langle \Theta_2, X \rangle + \langle \Theta_1, \mu_X \rangle + b + \varepsilon.\]
Therefore, it is not possible to distinguish between the term $\langle \Theta_1, \mu_X \rangle$ and the intercept term $b$, and consequently it is not possible to check whether $\Theta_1=0$ or not. Taking this into account, the hypothesis test will be restricted to check
\begin{equation}
  \left\{\begin{array}{ll} H_0: &\Theta_2=0 \\ H_1: &\Theta_2\neq 0 \end{array}\right. \label{eq.restrictedtest}
\end{equation}
on the basis of the available sample information.\\

Note that in this case, according to the relation between the operators and the regression parameter shown in \eqref{eq.relation}, $\Theta_2=0$ if, and only if, $\Delta(h)=0$ for all $h \in {\cal H}$. Consequently, the hypothesis test in \eqref{eq.restrictedtest} is equivalent to
\begin{equation}
  \left\{\begin{array}{ll} H_0: &\|\Delta\|'=0 \\ H_1: &\|\Delta\|'\neq 0 \end{array}\right. \label{eq.deltatest}
\end{equation}

\begin{rem}
It should be recalled that, in previous works $\mu_X$ is assumed to be equal $0$. Thus, the preceding reasoning leads to the fact that $\Theta_1$ cannot be estimated based on the information provided by $X$ (see, for instance, Cardot, Ferraty, Mas, and Sarda~(2003)). Consequently the hypothesis testing is also restricted to the one in the preceding equations. In addition in Cardot, Ferraty, Mas, and Sarda~(2003), it is also assumed for technical reasons that $\overline{{\cal R}(\Gamma)}$ is an infinite-dimensional space. On the contrary, this restriction is not imposed in the study here developed.
\end{rem}

\begin{rem}
Note that another usual assumption is that the intercept term vanishes. Although this is not common in most of situations, it should be noted that if $b=0$ and $X$ is not assumed to be centered (as in this work), then an interesting possibility appears: to check whether $\Theta_1=0$ or not by checking the nullity of the intercept term of the model, and thus to check the original hypothesis testing in \eqref{eq.test}. This open problem cannot be solved with the methodology employed in the current paper (or in the previous ones) because the idea is based on checking \eqref{eq.deltatest}, which is equivalent to the restricted test \eqref{eq.restrictedtest} but not to the unrestricted one in \eqref{eq.test}.
\end{rem}

\subsection{Testing procedure and asymptotic theory}

According to the relation between $\|\cdot\|'$ and $\|\cdot\|$, the dual norm of $\Delta \in {\cal H}'$ can be expressed equivalently in terms of the ${\cal H}$-valued random element $(X-\mu_X)(Y-\mu_Y)$ as follows
\[\|\Delta\|'=\|\langle E\left((X-\mu_X)(Y-\mu_Y)\right),\cdot\rangle\|'=\|E\left((X-\mu_X)(Y-\mu_Y)\right)\|.\]
Thus, based on an i.i.d. sample $\{(X_i,Y_i)\}_{i=1}^n$ drawn from $(X,Y)$,
\[D=\|E\left((X-\mu_X)(Y-\mu_Y)\right)\|=\|T\|\]
can be estimated in a natural way by means of its empirical counterpart $D_n=\|T_n\|$, where $T_n$ is the ${\cal H}$-valued random element given by 
\[T_n=\frac{1}{n}\sum_{i=1}^n (X_i-\overline{X})(Y_i-\overline{Y}),\]
where $\overline{X}$ and $\overline{Y}$ denote as usual the corresponding sample means. The next theorem establishes some basic properties of $T_n$.

\begin{thm} \label{th.conv}
Assuming that \eqref{eq.hlm} holds with $E(\varepsilon)=0$, $E(\varepsilon^2)=\sigma^2<\infty$ and $E(\|X\|^4)<\infty$, then
	\begin{enumerate}
	 \item $E(T_n)= E\left((X-\mu_X)(Y-\mu_Y)\right)(n-1)/n$
	 \item $T_n$ converges a.s.-$P$ to $E\left((X-\mu_X)(Y-\mu_Y)\right)$ as $n \rightarrow \infty$
	 \item $\sqrt{n} \left(T_n - E\left((X-\mu_X)(Y-\mu_Y)\right) \right)$ converges in law, as $n \rightarrow \infty$, to a centered Gaussian element $Z$ in ${\cal H}$ with covariance operator
\[\Gamma_Z(\cdot) = \sigma^2 \Gamma(\cdot) + E\left( (X-\mu_X)\langle X-\mu_X, \cdot \rangle \langle X-\mu_X, \Theta \rangle^2\right).\]
	\end{enumerate}
\end{thm}
\begin{proof}
Since $T_n$ can be equivalently expressed as
\[T_n = \frac{1}{n}\sum_{i=1}^n (X_i-\mu_X)(Y_i-\mu_Y) - (\overline{X} - \mu_X)(\overline{Y} - \mu_Y),\]
it is straightforward to check item~$1$. The a.s.-$P$ convergence is a direct application of the SLLN for separable Hilbert-valued random elements.\\

On the other hand, given that $E(\|(X-\mu_X)(Y-\mu_Y)\|^2)<\infty$, the convergence in law can be deduced by applying the CLT for separable Hilbert-valued random elements (see, for instance, Laha and Rohatgi~(1979)) together with Slutsky's Theorem. The concrete expression of the operator $\Gamma_Z$, that is, $\Gamma_Z = \Gamma_{(X-\mu_X)(Y-\mu_Y)} = \Gamma_{(X-\mu_X)\varepsilon} + \Gamma_{(X-\mu_X)\langle X-\mu_X, \Theta \rangle }$, can be obtained by simple computations.
\end{proof}

In order to simplify the notation, from now on, given any ${\cal H}$-valued random element $H$ with $E(\|H\|^2)<\infty$, $Z_H$ will denote a centered Gaussian element in ${\cal H}$ with covariance operator $\Gamma_H$.

\begin{cor} \label{cor.null}
Under the conditions of Theorem~\ref{th.conv}, if the null hypothesis $H_0: \|\Delta\|'=0$ is satisfied, then $\sqrt{n} T_n$ converges in law to $Z_{(X-\mu_X)\varepsilon}$ (with covariance operator $\sigma^2 \Gamma$), and consequently, $\|\sqrt{n} T_n\|$ converges in law to $\|Z_{(X-\mu_X)\varepsilon}\|$.
\end{cor}

In contrast to Theorem~1 in Cardot, Ferraty, Mas, and Sarda~(2003), the result in Corollary~\ref{cor.null} is established directly on the Hilbert space $\cal H$ instead of on its dual space. In addition, no assumption of centered $X$ random elements or null intercept term is necessary. Nevertheless these two assumptions could be easily removed in that paper in order to establish a dual result of Corollary~\ref{cor.null}.\\

Furthermore, in view of Corollary~\ref{cor.null}, the asymptotic null distribution of $\|\sqrt{n} T_n\|$ is not explicitly known. This is the reason why no further research on how to use in practice this statistic (or its dual one) for checking if $\Theta_2$ equals $0$ is carried out in Cardot, Ferraty, Mas, and Sarda~(2003). Instead, an alternative statistic that is used in the simulation section for comparative purposes is considered. Nevertheless, it is still possible to use $\|\sqrt{n} T_n\|$ as a core statistic in order to solve this test in practice by means of bootstrap techniques.\\

One natural way of using the asymptotic result of Corollary~\ref{cor.null} for solving the test under study is as follows. Consider a consistent (at least under $H_0$) estimator $\sigma^2_n$ of $\sigma^2$ (for instance, the sample variance of $Y$ could be used, or perhaps the one introduced by Cardot, Ferraty, Mas, and Sarda~(2003), provided that its theoretical behavior is analyzed). Then, according to Slutsky's Theorem $\|\sqrt{n} T_n\|/\sigma_n$ converges in law under $H_0$ to the norm of $Z_X$. As its covariance operator $\Gamma$ is unknown, it can be approximated by the empirical one $\Gamma_n$. And thus, $\|Z_X\|$ can be approximated by $\|Z_n\|$, being $Z_n$ a centered Gaussian element in ${\cal H}$ with covariance operator $\Gamma_n$. Of course, the distribution of $\|Z_n\|$ is still difficult to compute directly, nevertheless one can make use of the CLT and approximate its distribution by Monte Carlo method by the distribution of
\[\left\|\frac{1}{m}\sum_{i=1}^m (X_i^*-\overline{X})\right\|\]
for a large value of $m$, being $\{X_i^*\}_{i=1}^m$ i.i.d. random elements chosen at random from the fixed population $(X_1,\ldots,X_n)$. Obviously, this method is a precursor of the bootstrap procedures.\\

In order to complete the asymptotic study of the statistic $\|\sqrt{n} T_n\|$, its behavior under local alternatives is going to be analyzed. To this purpose, let us consider $\Theta \in {\cal H}$ so that $\|\Theta_2\|>0$, and given $\delta_n >0$ consider the modified random sample 
\[Y_i^n= \langle X_i, \frac{\delta_n}{\sqrt{n}} \Theta \rangle + b + \varepsilon_i,\]
for all $i \in \{1,\ldots,n\}$. Then, the null hypothesis is not verified. However, if $\delta_n/\sqrt{n} \rightarrow 0$, then $\|(\delta_n/\sqrt{n}) \Theta \| \rightarrow 0$, that is, $H_0$ is approached with ``speed'' $\delta_n/\sqrt{n}$. In these conditions,
\[E\left( (X_i-\mu_{X_i})(Y_i^n-\mu_{Y_i^n}) \right) = \frac{\delta_n}{\sqrt{n}} \Gamma(\Theta),\]
and thus the following theorem that establishes the behavior of the statistic under the considered local alternatives can be easily deduced.

\begin{thm}
Under the conditions of Theorem~\ref{th.conv} and with the above notation, if $\delta_n \rightarrow \infty$ and $\delta_n/\sqrt{n} \rightarrow 0$ as $n \rightarrow \infty$ then 
\[P\left( \left\|\frac{1}{\sqrt{n}} \sum_{i=1}^n \left(X_i - \overline{X}\right) \left( Y_i^n - \overline{Y^n} \right) \right\| \leq t\right) \rightarrow 0\]
as $n \rightarrow \infty$, for all $t \in \R$.
\end{thm}

\subsection{Bootstrap procedures}

The difficulty of using the previously proposed statistic to solve the hypothesis test by means of asymptotic procedures suggests the development of appropriated bootstrap techniques. The asymptotic consistency of a bootstrap approach is guaranteed if the associated bootstrap statistic converges in law to a non-degenerated distribution irrespectively of $H_0$ being satisfied or not. In addition, in order to ensure its asymptotic correctness, this limit distribution must coincide with the asymptotic one of the testing statistic provided that $H_0$ holds.\\

Consequently, the asymptotic limit established in Corollary~\ref{cor.null} plays a fundamental role for defining appropriate bootstrap statistics. In this way, recall that 
\[\frac{1}{\sqrt{n}}\sum_{i=1}^n \left( \big(X_i - \overline{X}\big)\big(Y_i - \overline{Y}\big) - E\big((X-\mu_X)(Y-\mu_Y)\big)\right)\]
converges in law to $Z_{(X-\mu_X)(Y-\mu_Y)}$, irrespectively of $H_0$ being satisfied or not and, in addition, if $H_0$ is satisfied then $\Gamma_{(X-\mu_X)(Y-\mu_Y)}=\sigma^2 \Gamma$. Thus, this is a natural statistic to be mimicked by a bootstrap one. Note that,  
\begin{equation}
\left(\frac{1}{n}\sum_{i=1}^n\big(Y_i - \overline{Y}\big)^2\right)\left(\frac{1}{\sqrt{n}}\sum_{i=1}^n \big(X_i-\mu_X\big) \right), \label{eq.stat2}
\end{equation}
converges in law to $(\sigma^2 + E(\langle X-\mu_X, \Theta\rangle^2)) Z_{X}$, whose covariance operator is $(\sigma^2 + E(\langle X-\mu_X, \Theta\rangle^2) \Gamma$. Particularly, when $H_0$ is satisfied, this operator reduces again to $\sigma^2 \Gamma$. Consequently, another possibility consists in mimicking this second statistic by means of a bootstrap one, improving the approximation suggested in the previous subsection. Note that the left term in the product in equation \eqref{eq.stat2} could be substituted by any other estimator under $H_0$ of $\sigma^2$ that converges to a finite constant if $H_0$ does not hold. Anyway, this second approximation could lead to worst results under the null hypothesis, because the possible dependency between $X$ and $\varepsilon$ is lost (as the resample would focus only on the $X$ information). \\

Two possibilities for mimicking the statistics which were above-mentioned are going to be explored, namely a ``naive'' paired bootstrap and a ``wild'' bootstrap approach. In order to achieve this goal, let $\{(X_i^*,Y_i^*)\}_{i=1}^n$ be a collection of i.i.d. random elements drawn at random from $(X_1,Y_1),\ldots,(X_n,Y_n)$, and let us consider the following ``naive'' paired bootstrap statistic
\[T_n^{N*}=\frac{1}{n}\sum_{i=1}^n \left( \big(X_i^* - \overline{X^*}\big)\big(Y_i^* - \overline{Y^*}\big) - \big(X_i - \overline{X}\big)\big(Y_i - \overline{Y}\big) \right).\]
In addition, let us consider $\sigma^2_n= (1/n)\sum_{i=1}^n(Y_i - \overline{Y})^2$ and $\sigma^{*2}_n= (1/n)\sum_{i=1}^n(Y^*_i - \overline{Y^*})^2$, the empirical estimator of $\sigma^2_Y$ under $H_0$ and its corresponding bootstrap version.\\

The asymptotic behavior of the ``naive'' bootstrap statistic will be analyzed through some results on bootstrapping general empirical measures obtained by Gin\'e and Zinn~(1990). It should be noted that the bootstrap results in that paper refer to empirical process indexed by a class of functions $\cal F$, that particularly extend to the bootstrap about the mean in separable Banach (and thus Hilbert) spaces. In order to establish this connection, it is enough to choose
\[{\cal F}= \{ f \in {\cal H}' | \| f \|'\leq 1\}\]
(see Gin\'e~(1997) and Kosorok~(2008), for a general overview of indexed empirical process). ${\cal F}$ is image admissible Suslin (considering the weak topology). In addition, $F(h)=\sup_{f \in {\cal F}}|f(h)|=\|h\|$ for all $h \in {\cal H}$ and thus $E(F^2(X))=E(\|X\|^2) < \infty$. \\\nowidow[3] 

Consider the bounded and linear (so continuous) operator $\delta$ from ${\cal H}$ to $l^{\infty}{\cal (F)}$ given by $\delta(h)(f)=\delta_h(f)=f(h)$ for all $h \in {\cal H}$ and all $f \in {\cal F}$ and denote by $R(\delta) \subset l^{\infty}{\cal (F)}$ its range. As $\|\delta(h)\|_\infty=\|h\|$ for all $h \in {\cal H}$ then, there exists $\delta^{-1}:R(\delta) \rightarrow {\cal H}$, so that $\delta^{-1}$ is continuous. In addition, as $R(\delta)$ is closed, Dugundji Theorem allows us to consider a continuous extension $\delta^{-1}:l^{\infty}{\cal (F)} \rightarrow {\cal H}$ (see for instance Kosorok~(2008), Lemma~6.16 and Theorem~10.9). Thus, following the typical empirical process notation, the empirical process $(1/\sqrt{n}) \sum_{i=1}^n (\delta_{X_i} - \mathbb{P})$ indexed in $\cal F$ is directly connected with $(1/\sqrt{n}) \sum_{i=1}^n (X_i - E(X))$ by means of the continuous mapping $\delta^{-1}$ and vice-versa. \\

Some consequences of this formulation applied to the work developed by Gin\'e and Zinn~(1990) lead to the results collected in following lemma.

\begin{lem} \label{le.1}
Let $\xi$ be a measurable mapping from a probabilistic space denoted by $(\Omega, \sigma, P)$ to a separable Hilbert space $({\cal H},\langle \cdot, \cdot \rangle)$ with corresponding norm $\|\cdot\|$ so that $E(\|\xi\|^2)<\infty$. Let $\{\xi_i\}_{i=1}^n$ be a sequence of i.i.d. random elements with the same distribution as $\xi$, and let $\{\xi^*_i\}_{i=1}^n$ be i.i.d. from $\{\xi_i\}_{i=1}^n$. Then
\begin{enumerate}
 \item $\sqrt{n}(\overline{\xi^*} - \overline{\xi})$ converges in law to $Z_{\xi}$ a.s.-$P$
 \item $\overline{\xi^*}$ converges in probability to $E(\xi)$ a.s.-$P$
 \item $\overline{\|\xi^*\|^2}$ converges in probability to $E(\|\xi\|^2)$ a.s.-$P$
\end{enumerate}
\end{lem}
\begin{proof}
To prove item~1 note that the CLT for separable Hilbert-valued random elements (see, for instance, Laha and Rohatgi~(1979)) together with the Continuous Mapping Theorem applied to $\delta$ guarantees that ${\cal F} \in \text{CLT}(P)$. Thus, Theorem~2.4 of Gin\'e and Zinn~(1990) ensures that $n^{1/2} (\hat{P}_n(w) - P_n(w))$ converges in law to a Gaussian process on $\cal F$, $G=\delta(Z_{\xi})$ a.s.-$P$. Consequently, by applying again the Continuous Mapping Theorem $\sqrt{n}(\overline{\xi^*} - \overline{\xi}) = \delta^{-1}(n^{1/2} (\hat{P}_n(w) - P_n(w)))$ converges in law to $Z_{\xi}=\delta^{-1}(G)$. \\

Items~2 and~3 can be checked in a similar way by applying Theorem~2.6 of Gin\'e and Zinn~(1990). Note that item~1 is also a direct consequence of Remark 2.5 of Gin\'e and Zinn~(1990); nevertheless it was proven based on Theorem~2.4 to illustrate the technique.  
\end{proof}

The following theorem establishes the asymptotic consistency and correctness of the ``naive'' bootstrap approach.

\begin{thm}
Under the conditions of Theorem~\ref{th.conv}, we have that $\sqrt{n}T_n^{N*}$ converges in law to\\ $Z_{(X-\mu_X)(Y-\mu_Y)}$ a.s.-$P$. In addition, $\sigma^{*2}_n$ converges in probability to $\sigma^2_{Y}=\sigma^2 + E(\langle X-\mu_X, \Theta\rangle^2)$ a.s.-$P$.
\end{thm}
\begin{proof}
First of all consider the bootstrap statistic 
\[S_n^*=\frac{1}{\sqrt{n}}\sum_{i=1}^n \left( \big(X_i^* - \mu_X\big)\big(Y_i^* - \mu_Y\big) - \overline{\big({X} - \mu_X\big)\big({Y} - \mu_Y\big)}\right)\]
and note that $\{\big(X_i^* - \mu_X\big)\big(Y_i^* - \mu_Y\big)\}_{i=1}^n$ are i.i.d. $\cal H$-valued random elements chosen at random from the ``bootstrap population'' $\{\big(X_i - \mu_X\big)\big(Y_i - \mu_Y\big)\}_{i=1}^n$. Then, item~1 in Lemma~\ref{le.1} guarantees that $S_n^*$ converges in law to $Z_{(X-\mu_X)(Y-\mu_Y)}$ a.s.-$P$.

On the other hand, $S_n^*$ equals $\sqrt{n}T_n^{N*}$ plus the following terms
\[-\frac{1}{\sqrt{n}}\sqrt{n}(\overline{X^*} - \overline{X})\sqrt{n}(\overline{Y^*} - \overline{Y}) + \sqrt{n}(\overline{X^*} - \overline{X})(\overline{Y^*} - \mu_Y) + (\overline{X^*} - \mu_{X})\sqrt{n}(\overline{Y^*} - \overline{Y}).\]
Items~1 and~2 in Lemma~\ref{le.1}, together with Slutsky's Theorem, ensure that these three terms converge in probability to $0$ a.s.-$P$, and consequently the convergence in law stated in the theorem is proven.\\

Finally, the convergence of $\sigma^{*2}_n$ holds in virtue of items~2 and~3 in Lemma~\ref{le.1}. 
\end{proof}

The ``naive'' bootstrap approach is described in the following algorithm.
\begin{alg}[Naive Bootstrap] \par\mbox{}

\begin{list}{\labelitemi}{\leftmargin=0.5cm}
 \item[Step 1.] Compute the value of the statistic $T_n$ (or the value $T_n/\sigma_n$). 
 \item[Step 2.] Draw $\{(X_i^*,Y_i^*)\}_{i=1}^n$, a sequence of i.i.d. random elements chosen at random from the initial sample $(X_1,Y_1),\ldots,(X_n,Y_n)$, and compute $a_n=\|T_n^{N*}\|$ (or $b_n=\|T_n^{N*}\|/\sigma^*_n$).
 \item[Step 3.] Repeat Step 2 a large number of times $B \in \N$ in order to obtain a sequence of values $\{a_n^l\}_{l=1}^B$ (or $\{b_n^l\}_{l=1}^B$).
 \item[Step 4.] Approximate the $p$-value of the test by the proportion of values in $\{a_n^l\}_{l=1}^B$ greater than or equal to $\|T_n\|$ (or by the proportion of values in $\{b_n^l\}_{l=1}^B$ greater than or equal to $\|T_n\|/\sigma_n$)
\end{list}
\end{alg}

Analogously, let $\{\varepsilon_i^*\}_{i=1}^n$ be i.i.d. centered real random variables so that $E\big((\varepsilon_i^*)^2\big)=1$ and \\$\int_0^{\infty} (P(|\varepsilon_1^*|>t)^{1/2})<\infty$ (to guarantee this last assumption, it is enough that $E\big((\varepsilon_i^*)^d\big)<\infty$ for certain $d>2$), and consider the ``wild'' bootstrap statistic 
\[T_n^{W*}=\frac{1}{n}\sum_{i=1}^n  \big(X_i - \overline{X}\big)\big(Y_i - \overline{Y}\big) \varepsilon_i^*.\]

In order to analyze the asymptotic behavior of the ``wild'' bootstrap statistic, the following lemma will be fundamental. It is a particularization of a result due to Ledoux, Talagrand and Zinn (cf. Gin\'e and Zinn~(1990), and Ledoux and Talagrand~(1988)). See also the Multiplier Central Limit Theorem in Kosorok~(2008) for the empirical process indexed by a class of measurable functions counterpart.

\begin{lem} \label{le.2}
Let $\xi$ be a measurable mapping from a probabilistic space denoted by $(\Omega, \sigma, P)$ to a separable Hilbert space $({\cal H},\langle \cdot, \cdot \rangle)$ with corresponding norm $\|\cdot\|$ so that $E(\|\xi\|^2)<\infty$. Let $\{\xi_i\}_{i=1}^n$ be a sequence of i.i.d. random elements with the same distribution as $\xi$, and let $\{W_i\}_{i=1}^n$ be a sequence of i.i.d. random variables (in the same probability space and independent of $\{\xi_i\}_{i=1}^n$) with $E(W_i)=0$ and $\int_0^{\infty} (P(|W_1|>t)^{1/2})<\infty$, then the following are equivalent
\begin{enumerate}
 \item $E(\|\xi\|^2)<\infty$ (and consequently $\sqrt{n}(\overline{\xi} - E({\xi}))$ converges in law to $Z_{\xi}$).
 \item For almost all $\omega \in \Omega$, $(1/\sqrt{n}) \sum_{i=1}^n W_i \xi_i(\omega)$ converges in law to $Z_{\xi}$.
\end{enumerate}
\end{lem}
 
As a consequence, the asymptotic consistency and correctness of the ``wild'' bootstrap approach is guaranteed by the following theorem.
\begin{thm}
Under the conditions of Theorem~\ref{th.conv}, we get that $\sqrt{n}T_n^{W*}$ converges in law to\\ $Z_{(X-\mu_X)(Y-\mu_Y)}$ a.s.-$P$.
\end{thm}
\begin{proof}
According to Lemma~\ref{le.2}, for almost all $\omega \in \Omega$,
\[S_n^*=\frac{1}{\sqrt{n}}\sum_{i=1}^n \big(X_i^w - \mu_X\big)\big(Y_i^w - \mu_Y\big)\varepsilon^*_i\]
converges in law to $Z_{(X-\mu_X)(Y-\mu_Y)}$. Moreover $(\overline{Y^w}-\mu_Y)$ and $(\overline{X^w}-\mu_X)$ converges to $0$ (by SLLN).\\

Finally note that, for almost all $\omega \in \Omega$,
\[\begin{split}S_n^* &= \sqrt{n}T_n^{W*} + (\overline{Y^w}-\mu_Y) \frac{1}{\sqrt{n}}\sum_{i=1}^n(X_i^w - \mu_X)\varepsilon^*_i\\
&+ (\overline{X^w}-\mu_X) \frac{1}{\sqrt{n}}\sum_{i=1}^n(Y_i^w - \mu_Y)\varepsilon^*_i
 - (\overline{X^w}-\mu_X)(\overline{Y^w}-\mu_Y) \frac{1}{\sqrt{n}}\sum_{i=1}^n\varepsilon^*_i.\end{split}\]
Lemma~\ref{le.2}, together with the SLLN above-mentioned, guarantees the convergence in probability to $0$ of the last three summands, and thus the result is reached in virtue of Slutsky's Theorem.
\end{proof}

The ``wild'' bootstrap approach proposed can be applied by means of the following algorithm.

\begin{alg}[Wild Bootstrap] \par\mbox{}
\begin{list}{\labelitemi}{\leftmargin=0.5cm}
 \item[Step 1.] Compute the value of the statistic $T_n$ (or the value $T_n/\sigma_n$). 
 \item[Step 2.] Draw $\{\varepsilon_i^*\}_{i=1}^n$ a sequence of i.i.d. random elements $\varepsilon$, and compute $a_n=\|T_n^{W*}\|$ 
(or $b_n=\|T_n^{W*}\|/\sigma^*_n$, in this case $\sigma^*_n$ is computed like in Step 2 of the Naive Bootstrap algorithm).
 \item[Step 3.] Repeat Step 2 a large number of times $B \in \N$ in order to obtain a sequence of values $\{a_n^l\}_{l=1}^B$ (or $\{b_n^l\}_{l=1}^B$).
 \item[Step 4.] Approximate the $p$-value of the test by the proportion of values in $\{a_n^l\}_{l=1}^B$ greater than or equal to $\|T_n\|$ (or by the proportion of values in $\{b_n^l\}_{l=1}^B$ greater than or equal to $\|T_n\|/\sigma_n$).
\end{list}
\end{alg}

\section{Bootstrap calibration vs. asymptotic theory}

For simplicity, suppose from now on that $b=0$ and $X$ of zero-mean in \eqref{eq.hlm}, that is, suppose that the regression model is given by
\[Y=\langle\Theta,X\rangle+\varepsilon.\]
Furthermore, $\Delta(h)=E\left(\langle X, h \rangle Y\right)$ and, analogously, $\Gamma(h)=E\left(\langle X, h \rangle X\right)$. In such case, if we assume that $\sum_{j=1}^{\infty}{(\Delta(v_j)/\lambda_j)^2}<+\infty$ and $\mathrm{Ker}(\Gamma)=\{0\}$, then
\[\Theta=\sum_{j=1}^{\infty}{\frac{\Delta(v_j)}{\lambda_j}v_j},\]
being $\{(\lambda_j,v_j)\}_{j\in\N}$ the eigenvalues and eigenfunctions of $\Gamma$ (see Cardot, Ferraty, and Sarda~(2003)).\\

A natural estimator for $\Theta$ is the FPCA estimator based on $k_n$ functional principal components given~by
\[\hat{\Theta}_{k_n}=\sum_{j=1}^{k_n}{\frac{\Delta_n(\hat{v}_j)}{\hat{\lambda}_j}\hat{v}_j},\]
where $\Delta_n$ is the empirical estimation of $\Delta$, that is, $\Delta_n(h)=(1/n)\sum_{i=1}^n{\langle X_i,h\rangle Y_i}$, and $\{(\hat{\lambda}_j,\hat{v}_j)\}_{j \in \N}$ are the eigenvalues and the eigenfunctions of $\Gamma_n$, the empirical estimator of $\Gamma$: $\Gamma_n(h)=(1/n)$ $\sum_{i=1}^n{\langle X_i,h\rangle X_i}$.\\

Different statistics can be used for testing the lack of dependence between $X$ and $Y$. Bearing in mind the expression \eqref{eq.restrictedtest}, one can think about using an estimator of $\|\Theta\|^2=\sum_{j=1}^{\infty}{(\Delta(v_j)/\lambda_j)^2}$ in order to test these hypotheses. In an alternative way, the expression \eqref{eq.deltatest} can be a motivation for different class of statistics based on the estimation of $\|\Delta\|'$.\\

One asymptotic distribution free based on the latter approach was given by Cardot, Ferraty, Mas, and Sarda~(2003). They proposed as test statistic
\begin{equation}
T_{1,n}=k_n^{-1/2}\left(\hat{\sigma}^{-2}||\sqrt{n}\Delta_n\hat{A}_n||'^2-k_n\right), \label{eq.t1}
\end{equation}
where $\hat{A}_n(\cdot)=\sum_{j=1}^{k_n}{\hat{\lambda}_j^{-1/2}\langle\cdot,\hat{v}_j\rangle\hat{v}_j}$ and $\hat{\sigma}^2$ is a consistent estimator of $\sigma^2$. Cardot, Ferraty, Mas, and Sarda~(2003) showed that, under $H_0$, $T_{1,n}$ converges in distribution to a centered Gaussian variable with variance equal to 2. Hence, $H_0$ is rejected if $|T_{1,n}|>\sqrt{2}z_{1-\alpha/2}$ ($z_{\alpha}$ the $\alpha$-quantile of a $\mathcal{N}(0,1)$), and accepted otherwise. Besides, Cardot, Ferraty, Mas, and Sarda~(2003) also proposed another calibration of the statistic distribution based on a permutation mechanism.\\

On the other hand, taking into account that $||\Theta||^2=\sum_{j=1}^{\infty}{(\Delta(v_j)/\lambda_j)^2}$, one can use the statistic
\begin{equation}
T_{2,n}=\sum_{j=1}^{k_n}{\left(\frac{\Delta_n(\hat{v}_j)}{\hat{\lambda}_j}\right)^2}, \label{eq.t2}
\end{equation}
which limit distribution is not known.\\

Finally, a natural competitive statistic is the one proposed throughout Section~2.3
\begin{equation}
T_{3,n}=\left\|\frac{1}{n}\sum_{i=1}^{n}{(X_i-\bar{X})(Y_i-\bar{Y})}\right\|, \label{eq.t3}
\end{equation}
which we will denote by ``$F$-test'' from now on since it is the natural generalization of the well-known $F$-test in the finite-dimensional context. Another possibility is to consider the studentized version of \eqref{eq.t3}
\begin{equation}
T_{3s,n}=\frac{1}{\hat{\sigma}}\left\|\frac{1}{n}\sum_{i=1}^{n}{(X_i-\bar{X})(Y_i-\bar{Y})}\right\|, \label{eq.t3s}
\end{equation}
where $\hat{\sigma}^2$ is the empirical estimation of $\sigma^2$.\\

In general, for the statistics such as \eqref{eq.t1}, \eqref{eq.t2}, \eqref{eq.t3} and \eqref{eq.t3s}, the calibration of the distribution can be obtained by using bootstrap. Furthermore, in the previous section, ``naive'' and ``wild'' bootstrap were shown to be consistent for the $F$-test, that is, the distribution of $T_{3,n}$ and $T_{3s,n}$ can be approximated by their corresponding bootstrap distribution, and $H_0$ can be rejected when the statistic value does not belong to the interval defined for the bootstrap acceptation region of confidence $1-\alpha$. The same calibration bootstrap can be applied to the tests based on $T_{1,n}$ and $T_{2,n}$, although the consistence of the bootstrap procedure in this cases have not been proved in this work.

\section{Simulation and real data applications}

In this section a simulation study and an application to a real dataset illustrate the performance of the asymptotic approach and the bootstrap calibration from a practical point of view.

\subsection{Simulation study}

We have simulated $ns=500$ samples, each being composed of $n\in\{50,100\}$ observations from the functional linear model $Y=\langle\Theta,X\rangle+\varepsilon$, being $X$ a Brownian motion and $\varepsilon\sim \mathcal{N}(0,\sigma^2)$ with signal-to-noise ratio $r=\sigma/\sqrt{\mathbb{E}(\langle X,\Theta\rangle^2)}\in\{0.5,1,2\}$.\\

Under $H_0$, we have considered the model parameter $\Theta_0(t)=0$, $t\in[0,1]$, whereas under $H_1$, the selected model parameter was $\Theta_1(t)=\sin(2\pi t^3)^3$, $t\in[0,1]$. Furthermore, under $H_0$ we have chosen $\sigma=1$, while in the alternative $H_1$ we assigned the three different values that were commented before. Let us remark that both $X$ and $\Theta$ were discretized to 100 equidistant design points.\\

We have selected the statistical tests which were introduced in the previous section: \eqref{eq.t1}, \eqref{eq.t2}, \eqref{eq.t3} and \eqref{eq.t3s}. For \eqref{eq.t1}, three distribution approximations were considered: the asymptotic approach ($\mathcal{N}(0,2)$) and the following two bootstrap calibrations
\begin{eqnarray*}
 T_{1,n}^{*(a)} & = & \frac{1}{\sqrt{k_n}}\left(\frac{n}{(\hat{\sigma}^{*})^2}\sum_{j=1}^{k_n}{\frac{\left(\Delta_n^{*}(\hat{v}_j)\right)^2} {\hat{\lambda}_j}}-k_n\right), \\
 T_{1,n}^{*(b)} & = & \frac{1}{\sqrt{k_n}}\left(\frac{n}{\hat{\sigma}^2}\sum_{j=1}^{k_n}{\frac{\left(\Delta_n^{*}(\hat{v}_j)\right)^2} {\hat{\lambda}_j}}-k_n\right).
\end{eqnarray*}
The difference between the two proposed bootstrap approximations is that in the former the estimation of $\sigma^2$ is also bootstrapped in each iteration. On the other hand, for \eqref{eq.t2}, \eqref{eq.t3} and \eqref{eq.t3s}, only the bootstrap approaches were computed
\begin{eqnarray*}
 T_{2,n}^{*} & = & \sum_{j=1}^{k_n}{\left(\frac{\Delta_n^{*}(\hat{v}_j)}{\hat{\lambda}_j}\right)^2}, \\
 T_{3,n}^{*} & = & \left\|\frac{1}{n}\sum_{i=1}^{n}{(X_i-\bar{X})(Y_i-\bar{Y})\varepsilon_i^*}\right\|, \\
 T_{3s,n}^{*} & = & \frac{1}{\hat{\sigma}^{*}}\left\|\frac{1}{n}\sum_{i=1}^{n}{(X_i-\bar{X})(Y_i-\bar{Y})\varepsilon_i^*}\right\|.
\end{eqnarray*}

For this simulation study, we have used the ``wild'' bootstrap algorithm introduced in Section~2.4 for the $F$-test and its studentized version, and the following adaptation of this consistent ``wild'' bootstrap for $T_{1,n}$ and $T_{2,n}$.

\begin{alg}[Wild Bootstrap] \par\mbox{}
\begin{list}{\labelitemi}{\leftmargin=0.5cm}
 \item[Step 1.] Compute the value of the statistic $T_{1,n}$ (or the value $T_{2,n}$). 
 \item[Step 2.] Draw $\{\varepsilon_i^*\}_{i=1}^n$ a sequence of i.i.d. random elements $\varepsilon$, and define $Y_i^{*}=Y_i\varepsilon_i^*$ for all $i=1,\ldots,n$.
 \item[Step 3.] Build $\Delta_n^{*}(\cdot)=n^{-1}\sum_{i=1}^{n}{\langle X_i,\cdot\rangle Y_i^{*}}$ and compute $a_n=|T_{1,n}^{*}|$ (or $b_n=|T_{2,n}^{*}|$).
 \item[Step 4.] Repeat Steps 2 and 3 a large number of times $B \in \N$ in order to obtain a sequence of values $\{a_n^l\}_{l=1}^B$ (or $\{b_n^l\}_{l=1}^B$).
 \item[Step 5.] Approximate the $p$-value of the test by the proportion of values in $\{a_n^l\}_{l=1}^B$ greater than or equal to $|T_{1,n}|$ (or by the proportion of values in $\{b_n^l\}_{l=1}^B$ greater than or equal to $|T_{2,n}|$).
\end{list}
\end{alg}
Let us indicate that $1,000$ bootstrap iterations were done in each simulation.\\

Due to $k_n$ and $\alpha$ must be fixed to run the procedure, the study was repeated with different numbers of principal components involved ($k_n\in\{1,\ldots,20\}$) and confidence levels ($\alpha\in\{0.2,0.1,0.05,0.01\}$). Nevertheless, in order to simplify the reading, the information collected in the following tables corresponds to only three of the values of $k_n$ which were analyzed: $k_n=5$, $k_n=10$ and $k_n=20$.\\

Table~\ref{tab:level} on page~\pageref{tab:level} displays the sizes of the test statistics obtained in the simulation study. For $T_{1,n}$, it can be highlighted that bootstrap approaches have closer sizes to the theoretical $\alpha$ than the asymptotic approximation for $T_{1,n}$, mainly when $k_n$ is small. If we compare the performance of the two bootstrap procedures proposed, it seems that if $\sigma^2$ is bootstrapped ($T_{1,n}^{*(a)}$) the results are better than if the same estimation of the variance is considered in all the bootstrap replications ($T_{1,n}^{*(b)}$) above all when $k_n$ is large. As far as $T_{2,n}$ is concerned, the estimated levels are quite near to the nominal ones, being $k_n=20$ the case in which they are farther from the theoretical $\alpha$. Finally, it must be remarked that the $F$-test and its studentized versions also get good results in terms of test levels, which are slightly closer to $\alpha$ when one uses the bootstrap distribution of $T_{3s,n}^{*}$ to approximate the distribution of the statistic.\\

\begin{table}[H] \centering{
\begin{tabular}{|ccrrrrrrrrrrrrrr|}\hline
\multirow{3}{*}{$n$} & \multirow{3}{*}{$\alpha$} & \multicolumn{3}{c}{$\mathcal{N}(0,2)$} & \multicolumn{3}{c}{$T_{1,n}^{*(a)}$} & \multicolumn{3}{c}{$T_{1,n}^{*(b)}$} & \multicolumn{3}{c}{$T_{2,n}^{*}$} & \multirow{3}{*}{$T_{3,n}^{*}$} & \multirow{3}{*}{$T_{3s,n}^{*}$} \\\cline{3-14}
 &  & \multicolumn{3}{c}{$k_n$} & \multicolumn{3}{c}{$k_n$} & \multicolumn{3}{c}{$k_n$} & \multicolumn{3}{c}{$k_n$} &  &  \\
 &  & \multicolumn{1}{c}{5} & \multicolumn{1}{c}{10} & \multicolumn{1}{c}{20} & \multicolumn{1}{c}{5} & \multicolumn{1}{c}{10} & \multicolumn{1}{c}{20} & \multicolumn{1}{c}{5} & \multicolumn{1}{c}{10} & \multicolumn{1}{c}{20} & \multicolumn{1}{c}{5} & \multicolumn{1}{c}{10} & \multicolumn{1}{c}{20} &  &  \\\hline
\multirow{4}{*}{50}	& 20\% & 19.4 & 17.6 & 16.0 & 21.4 & 21.6 & 20.0 & 21.6 & 19.0 & 15.2 & 19.8 & 20.8 & 18.4 & 21.6 & 20.8 \\
		& 10\% & 10.8 & 10.4 &  8.2 &  9.0 & 10.8 & 10.6 &  8.0 &  7.2 &  3.2 &  8.6 &  7.2 &  7.2 & 11.8 & 11.2 \\
		&  5\% &  8.2 &  7.0 &  4.4 &  5.0 &  4.0 &  4.6 &  5.0 &  2.4 &  0.0 &  4.0 &  3.2 &  3.0 &  6.0 &  6.2 \\
		&  1\% &  4.8 &  4.2 &  2.2 &  1.2 &  0.4 &  0.0 &  0.6 &  0.0 &  0.0 &  0.2 &  0.6 &  0.4 &  0.6 &  1.2 \\\hline
\multirow{4}{*}{100} & 20\% & 15.0 & 19.4 & 20.0 & 20.8 & 21.0 & 19.0 & 21.0 & 20.8 & 18.0 & 21.4 & 19.4 & 17.6 & 21.6 & 21.2 \\
		& 10\% &  8.6 &  9.6 &  9.0 & 11.8 & 10.8 & 10.4 & 10.4 &  9.6 &  6.2 &  9.8 &  8.8 &  7.0 & 11.6 & 11.8 \\
		&  5\% &  5.6 &  5.2 &  4.0 &  4.4 &  4.6 &  3.6 &  3.6 &  3.4 &  2.2 &  4.6 &  5.2 &  2.8 &  5.6 &  5.6 \\
		&  1\% &  2.6 &  2.4 &  1.2 &  1.4 &  1.2 &  0.8 &  1.2 &  0.6 &  0.2 &  1.0 &  0.6 &  0.8 &  0.4 &  0.4 \\\hline
\end{tabular}}
\caption{\small Comparison of the estimated levels for $T_{1,n}$ (using the asymptotic distribution $\mathcal{N}(0,2)$ and the bootstrap distributions of $T_{1,n}^{*(a)}$ and $T_{1,n}^{*(b)}$),  $T_{2,n}$ (using the bootstrap distribution of $T_{2,n}^{*}$), $T_{3,n}$ (using the bootstrap distribution of $T_{3,n}^{*}$) and its studentized version, $T_{3s,n}$ (using the bootstrap distribution of $T_{3s,n}^{*}$).}\label{tab:level}
\end{table}
\begin{table}[H] \centering{
\begin{tabular}{|ccrrrrrrrrrrrrrr|}\hline
\multirow{3}{*}{$n$} & \multirow{3}{*}{$\alpha$} & \multicolumn{3}{c}{$\mathcal{N}(0,2)$} & \multicolumn{3}{c}{$T_{1,n}^{*(a)}$} & \multicolumn{3}{c}{$T_{1,n}^{*(b)}$} & \multicolumn{3}{c}{$T_{2,n}^{*}$} & \multirow{3}{*}{$T_{3,n}^{*}$} & \multirow{3}{*}{$T_{3s,n}^{*}$} \\\cline{3-14}
 &  & \multicolumn{3}{c}{$k_n$} & \multicolumn{3}{c}{$k_n$} & \multicolumn{3}{c}{$k_n$} & \multicolumn{3}{c}{$k_n$} &  &  \\
 &  & \multicolumn{1}{c}{5} & \multicolumn{1}{c}{10} & \multicolumn{1}{c}{20} & \multicolumn{1}{c}{5} & \multicolumn{1}{c}{10} & \multicolumn{1}{c}{20} & \multicolumn{1}{c}{5} & \multicolumn{1}{c}{10} & \multicolumn{1}{c}{20} & \multicolumn{1}{c}{5} & \multicolumn{1}{c}{10} & \multicolumn{1}{c}{20} &  &  \\\hline
\multirow{4}{*}{50}	& 20\% & 100.0 & 100.0 & 100.0 & 100.0 & 100.0 & 100.0 & 100.0 & 100.0 & 100.0 &  88.8 &  0.0 & 0.0 & 100.0 & 100.0 \\
		& 10\% & 100.0 & 100.0 & 100.0 & 100.0 & 100.0 & 100.0 & 100.0 & 100.0 & 100.0 &  60.8 &  0.0 & 0.0 & 100.0 & 100.0 \\
		&  5\% & 100.0 & 100.0 & 100.0 & 100.0 & 100.0 & 100.0 & 100.0 & 100.0 &  99.0 &  32.2 &  0.0 & 0.0 & 100.0 & 100.0 \\
		&  1\% & 100.0 & 100.0 & 100.0 & 100.0 & 100.0 & 100.0 & 100.0 &  99.4 &  51.4 &   3.4 &  0.0 & 0.0 &  99.4 & 100.0 \\\hline
\multirow{4}{*}{100} & 20\% & 100.0 & 100.0 & 100.0 & 100.0 & 100.0 & 100.0 & 100.0 & 100.0 & 100.0 & 100.0 &  1.0 & 0.0 & 100.0 & 100.0 \\
		& 10\% & 100.0 & 100.0 & 100.0 & 100.0 & 100.0 & 100.0 & 100.0 & 100.0 & 100.0 & 100.0 &  0.0 & 0.0 & 100.0 & 100.0 \\
		&  5\% & 100.0 & 100.0 & 100.0 & 100.0 & 100.0 & 100.0 & 100.0 & 100.0 & 100.0 &  98.4 &  0.0 & 0.0 & 100.0 & 100.0 \\
		&  1\% & 100.0 & 100.0 & 100.0 & 100.0 & 100.0 & 100.0 & 100.0 & 100.0 & 100.0 &  70.0 &  0.0 & 0.0 & 100.0 & 100.0 \\ \hline
\end{tabular}}
\caption{\small For $r=0.5$, comparison of the empirical power for $T_{1,n}$ (using the asymptotic distribution $\mathcal{N}(0,2)$ and the bootstrap distributions of $T_{1,n}^{*(a)}$ and $T_{1,n}^{*(b)}$),  $T_{2,n}$ (using the bootstrap distribution of $T_{2,n}^{*}$), $T_{3,n}$ (using the bootstrap distribution of $T_{3,n}^{*}$) and its studentized version, $T_{3s,n}$ (using the bootstrap distribution of $T_{3s,n}^{*}$).}\label{tab:power.0.5}
\end{table}

On the other hand, Table~\ref{tab:power.0.5} on page~\pageref{tab:power.0.5}, Table~\ref{tab:power.1} on page~\pageref{tab:power.1}, and Table~\ref{tab:power.2} on page~\pageref{tab:power.2} show the empirical power obtained with the different procedures for each considered signal-to-noise ratio $r$. In terms of power, when $r=0.5$ the results for all the methods are similar, except for $T_{2,n}$ for which the empirical power decreases drastically, above all when $k_n$ increases (this effect is also observed for $r=1$ and $r=2$). This fact seems to be due to the construction of $T_{2,n}$ since this test statistic is the only one which does not involve the estimation of $\sigma^2$. In addition, the power of $T_{1,n}$ also falls\nopagebreak[4] abruptly when $T_{1,n}^{*(b)}$ is considered, $n$ is small and $k_n$ is very large.\\
 
A similar situation can be observed when $r=1$ and $r=2$. In the latter it can be seen that the empirical power is smaller for all the methods in general, being obtained an important loss of power when the sample is small ($n=50$), and $k_n$ increases and/or $\alpha$ decreases (see Table~\ref{tab:power.2} on page~\pageref{tab:power.2}). Furthermore, in this case, it can be seen that the empirical power relies heavily on the selected $k_n$ value. Hence, the advantage of using $T_{3,n}$ or $T_{3s,n}$ is that they do not require the selection of any parameter and they are competitive in terms of power. Nevertheless, it also seems that an adequate $k_n$ selection can make $T_{1,n}$ obtain larger empirical power than $T_{3,n}$ or $T_{3s,n}$ in some cases.\\

\begin{table}[H] \centering{
\begin{tabular}{|ccrrrrrrrrrrrrrr|}\hline
\multirow{3}{*}{$n$} & \multirow{3}{*}{$\alpha$} & \multicolumn{3}{c}{$\mathcal{N}(0,2)$} & \multicolumn{3}{c}{$T_{1,n}^{*(a)}$} & \multicolumn{3}{c}{$T_{1,n}^{*(b)}$} & \multicolumn{3}{c}{$T_{2,n}^{*}$} & \multirow{3}{*}{$T_{3,n}^{*}$} & \multirow{3}{*}{$T_{3s,n}^{*}$} \\\cline{3-14}
 &  & \multicolumn{3}{c}{$k_n$} & \multicolumn{3}{c}{$k_n$} & \multicolumn{3}{c}{$k_n$} & \multicolumn{3}{c}{$k_n$} &  &  \\
 &  & \multicolumn{1}{c}{5} & \multicolumn{1}{c}{10} & \multicolumn{1}{c}{20} & \multicolumn{1}{c}{5} & \multicolumn{1}{c}{10} & \multicolumn{1}{c}{20} & \multicolumn{1}{c}{5} & \multicolumn{1}{c}{10} & \multicolumn{1}{c}{20} & \multicolumn{1}{c}{5} & \multicolumn{1}{c}{10} & \multicolumn{1}{c}{20} &  &  \\\hline
\multirow{4}{*}{50}	& 20\% & 100.0 & 100.0 & 100.0 & 100.0 & 100.0 & 100.0 & 100.0 & 100.0 &  98.2 &  66.6 &  3.6 & 0.2 & 100.0 & 100.0 \\
		& 10\% & 100.0 & 100.0 & 100.0 & 100.0 & 100.0 &  99.8 & 100.0 &  99.8 &  89.6 &  33.6 &  0.8 & 0.0 & 100.0 & 100.0 \\
		&  5\% & 100.0 & 100.0 &  99.8 & 100.0 & 100.0 &  99.6 & 100.0 &  99.0 &  59.6 &  16.6 &  0.2 & 0.0 &  99.2 &  99.2 \\
		&  1\% & 100.0 & 100.0 &  99.6 &  99.6 &  97.6 &  94.6 &  95.2 &  67.6 &   2.6 &   2.2 &  0.0 & 0.0 &  87.8 &  92.4 \\\hline
\multirow{4}{*}{100} & 20\% & 100.0 & 100.0 & 100.0 & 100.0 & 100.0 & 100.0 & 100.0 & 100.0 & 100.0 &  97.0 &  7.8 & 0.0 & 100.0 & 100.0 \\
		& 10\% & 100.0 & 100.0 & 100.0 & 100.0 & 100.0 & 100.0 & 100.0 & 100.0 & 100.0 &  86.4 &  2.2 & 0.0 & 100.0 & 100.0 \\
		&  5\% & 100.0 & 100.0 & 100.0 & 100.0 & 100.0 & 100.0 & 100.0 & 100.0 & 100.0 &  67.8 &  1.0 & 0.0 & 100.0 & 100.0 \\
		&  1\% & 100.0 & 100.0 & 100.0 & 100.0 & 100.0 & 100.0 & 100.0 & 100.0 &  99.8 &  21.6 &  0.2 & 0.0 & 100.0 & 100.0 \\ \hline
\end{tabular}}
\caption{\small For $r=1$, comparison of the empirical power for $T_{1,n}$ (using the asymptotic distribution $\mathcal{N}(0,2)$ and the bootstrap distributions of $T_{1,n}^{*(a)}$ and $T_{1,n}^{*(b)}$),  $T_{2,n}$ (using the bootstrap distribution of $T_{2,n}^{*}$), $T_{3,n}$ (using the bootstrap distribution of $T_{3,n}^{*}$) and its studentized version, $T_{3s,n}$ (using the bootstrap distribution of $T_{3s,n}^{*}$).}\label{tab:power.1}
\end{table}
\begin{table}[H] \centering{
\begin{tabular}{|ccrrrrrrrrrrrrrr|}\hline
\multirow{3}{*}{$n$} & \multirow{3}{*}{$\alpha$} & \multicolumn{3}{c}{$\mathcal{N}(0,2)$} & \multicolumn{3}{c}{$T_{1,n}^{*(a)}$} & \multicolumn{3}{c}{$T_{1,n}^{*(b)}$} & \multicolumn{3}{c}{$T_{2,n}^{*}$} & \multirow{3}{*}{$T_{3,n}^{*}$} & \multirow{3}{*}{$T_{3s,n}^{*}$} \\\cline{3-14}
 &  & \multicolumn{3}{c}{$k_n$} & \multicolumn{3}{c}{$k_n$} & \multicolumn{3}{c}{$k_n$} & \multicolumn{3}{c}{$k_n$} &  &  \\
 &  & \multicolumn{1}{c}{5} & \multicolumn{1}{c}{10} & \multicolumn{1}{c}{20} & \multicolumn{1}{c}{5} & \multicolumn{1}{c}{10} & \multicolumn{1}{c}{20} & \multicolumn{1}{c}{5} & \multicolumn{1}{c}{10} & \multicolumn{1}{c}{20} & \multicolumn{1}{c}{5} & \multicolumn{1}{c}{10} & \multicolumn{1}{c}{20} &  &  \\\hline
\multirow{4}{*}{50}	& 20\% &  85.4 &  75.6 &  66.8 &  89.0 &  81.2 &  77.2 &  89.0 &  76.8 &  51.4 &  34.0 & 11.8 & 7.2 &  90.4 &  89.8 \\
		& 10\% &  80.0 &  68.6 &  56.4 &  79.4 &  68.6 &  59.4 &  76.4 &  57.4 &  20.2 &  16.6 &  4.0 & 2.4 &  79.0 &  79.0 \\
		&  5\% &  74.4 &  62.2 &  48.4 &  67.4 &  51.6 &  43.6 &  60.8 &  37.8 &   6.2 &  10.4 &  1.0 & 0.4 &  67.8 &  67.2 \\
		&  1\% &  67.4 &  51.4 &  35.6 &  40.0 &  26.4 &  20.2 &  25.4 &   6.0 &   0.0 &   0.8 &  0.0 & 0.0 &  34.4 &  39.0 \\\hline
\multirow{4}{*}{100} & 20\% &  99.8 &  98.8 &  94.6 & 100.0 &  99.8 &  98.0 & 100.0 &  99.2 &  94.2 &  60.0 & 14.6 & 7.6 &  99.8 &  99.8 \\
		& 10\% &  99.6 &  96.6 &  91.2 &  99.6 &  97.2 &  93.6 &  99.6 &  96.0 &  82.4 &  34.2 &  6.2 & 2.0 &  97.8 &  97.4 \\
		&  5\% &  99.6 &  95.6 &  85.8 &  97.8 &  94.0 &  85.8 &  97.2 &  90.4 &  64.6 &  18.0 &  2.8 & 0.4 &  94.4 &  94.4 \\
		&  1\% &  97.6 &  91.4 &  75.4 &  88.2 &  76.4 &  64.0 &  85.2 &  63.4 &  26.2 &   2.2 &  0.8 & 0.0 &  79.2 &  82.4 \\ \hline
\end{tabular}}
\caption{\small For $r=2$, comparison of the empirical power for $T_{1,n}$ (using the asymptotic distribution $\mathcal{N}(0,2)$ and the bootstrap distributions of $T_{1,n}^{*(a)}$ and $T_{1,n}^{*(b)}$),  $T_{2,n}$ (using the bootstrap distribution of $T_{2,n}^{*}$), $T_{3,n}$ (using the bootstrap distribution of $T_{3,n}^{*}$) and its studentized version, $T_{3s,n}$ (using the bootstrap distribution of $T_{3s,n}^{*}$).}\label{tab:power.2}
\end{table}

\subsection{Data application}

For the real data application, we have obtained concentrations of hourly averaged NO$_x$ in the neighborhood of a power station belonging to ENDESA, located in As Pontes in the Northwest of Spain. During unfavorable meteorological conditions, NO$_x$ levels can quickly rise and cause an air-quality episode. The aim is to forecast NO$_x$ with half an hour horizon to allow the power plant staff to avoid NO$_x$ concentrations reaching the limit values fixed by the current environmental legislation. This fact implies that it is necessary to estimate properly the regression model which defines the relationship between the observed NO$_x$ concentration in the last minutes ($X$) and the NO$_x$ concentration with half an hour horizon ($Y$). For that, a first step is to determine if there exists a linear dependence between $X$ and $Y$.\\

Therefore, we have built a sample where each curve $X$ corresponds to 240 consecutive minutal values of hourly averaged NO$_x$ concentration, and the response $Y$ corresponds to the NO$_x$ value half an hour ahead (from Jan 2007 to Dec 2009). Applying the tests for dependence to the dataset, the null hypothesis is rejected in all cases (thus, there is a linear relationship between the variables), except for $T_{2,n}$ when $k_n$ is large (see Table~\ref{tab:application} on page~\pageref{tab:application}). Nevertheless, as we have commented in the simulation study, this test statistic does not take into account the variance term and its power is clearly lower than the power of the other tests.\\

\begin{table}[H] \centering{
\begin{tabular}{|rrrrrrrrrrrrrr|}\hline
\multicolumn{3}{|c}{$\mathcal{N}(0,2)$} & \multicolumn{3}{c}{$T_{1,n}^{*(a)}$} & \multicolumn{3}{c}{$T_{1,n}^{*(b)}$} & \multicolumn{3}{c}{$T_{2,n}^{*}$} & \multirow{3}{*}{$T_{3,n}^{*}$} & \multirow{3}{*}{$T_{3s,n}^{*}$} \\\cline{1-12}
\multicolumn{3}{|c}{$k_n$} & \multicolumn{3}{c}{$k_n$} & \multicolumn{3}{c}{$k_n$} & \multicolumn{3}{c}{$k_n$} &  &  \\
\multicolumn{1}{|c}{1} & \multicolumn{1}{c}{5} & \multicolumn{1}{c}{10} & \multicolumn{1}{c}{1} & \multicolumn{1}{c}{5} & \multicolumn{1}{c}{10} & \multicolumn{1}{c}{1} & \multicolumn{1}{c}{5} & \multicolumn{1}{c}{10} & \multicolumn{1}{c}{1} & \multicolumn{1}{c}{5} & \multicolumn{1}{c}{10} &  &  \\\hline
0.000 & 0.000 & 0.000 & 0.000 & 0.000 & 0.000 & 0.000 & 0.000 & 0.000 & 0.000 & 0.002 & 0.011 & 0.000 & 0.000 \\\hline
\end{tabular}}
\caption{\small Real data application. $p$-values for $T_{1,n}$ (using the asymptotic distribution $\mathcal{N}(0,2)$ and the bootstrap distributions of $T_{1,n}^{*(a)}$ and $T_{1,n}^{*(b)}$),  $T_{2,n}$ (using the bootstrap distribution of $T_{2,n}^{*}$), $T_{3,n}$ (using the bootstrap distribution of $T_{3,n}^{*}$) and its studentized version, $T_{3s,n}$ (using the bootstrap distribution of $T_{3s,n}^{*}$).}\label{tab:application}
\end{table}

\section{Final comments}

The proposed bootstrap methods seems to give test sizes closer to the nominal ones than the tests based on the asymptotic distributions. In terms of power, the statistic tests which include a consistent estimation of the error variance $\sigma^2$ are better that the tests which do not take it into account. Furthermore, in all the cases, a suitable choice of $k_n$ seems to be quite important, and currently it is still an open question.\\

Besides of the optimal $k_n$ selection, other issues related to these dependence tests require further research, such as their extension to functional linear models with functional response. On the other hand, and in addition to the natural usefulness of this test, if would be interesting to combine it with the functional ANOVA test (see Cuevas, Febrero, and Fraiman~(2004), and Gonz\'alez-Rodr\'iguez, Colubi, and Gil~(2012)) in order to develop an ANCOVA test in this context.

\section*{Acknowledgements}

The work of the first and third authors was supported by Ministerio de Ciencia e Innovaci\'on (project MTM2008-03010), and by Conseller\'ia de Innovaci\'on e Industria (project PGIDIT07PXIB207031PR), and Conseller\'ia de Econom\'ia e Industria (project 10MDS207015PR), Xunta de Galicia. The work of the second author was supported by Ministerio de Ciencia e Innovaci\'on (project MTM2009-09440-C0202) and by the COST Action IC0702. The work of the fourth author was supported by Ministerio de Educaci\'on (FPU grant AP2010-0957).

\end{document}